\pdfoutput=1

\documentclass[sigconf,authorversion,nonacm,screen]{acmart}

\setcopyright{cc}
\setcctype{by-nc-nd}
\copyrightyear{2023}

\usepackage[acronym]{glossaries}
\glsdisablehyper

\newacronym{acr:AOS}{AOS}{Airborne Optical Sectioning}
\newacronym{acr:PSO}{PSO}{Particle Swarm Optimization}
\newacronym{acr:APSO}{APSO}{Adaptive PSO}
\newacronym{acr:CPSO}{CPSO}{Charged PSO}

\usepackage{algorithm}
\usepackage{algpseudocode}

\usepackage[noabbrev]{cleveref}

\usepackage[bb=dsserif]{mathalpha}

\usepackage{enumitem}
\setitemize{leftmargin=*}

\usepackage{extdash}

\usepackage{tabularx}

\begin{document}

\author{Julia Pöschl}
\affiliation{
    \institution{Johannes Kepler University}
    \city{Linz}
    \country{Austria}
}

\title[Adaptive Particle Swarm Optimization for through-foliage target detection with drone swarms]
{Adaptive Particle Swarm Optimization \\ for through-foliage target detection with drone swarms}

\begin{abstract}
  This work contributes to efforts on autonomously detecting a vegetation\Hyphdash occluded target by airborne observers. 
  It investigates and enhances previous work on a \gls{acr:PSO} strategy for \gls{acr:AOS} drone swarms. 
  First, it identifies two issues with that method and proposes to resolve them by a leader stabilization for its 
  scattering and projection-based line positions for its default scanning pattern. 
  Second, it connects that method to other \gls{acr:PSO} variants and presents a new adaptive \gls{acr:PSO} strategy for \gls{acr:AOS} drone swarms that draws on the ideas of \gls{acr:APSO}. 
\end{abstract}

\maketitle
\glsresetall

\section{Introduction}\label{sec:intro}

Drones empower us to cover even otherwise inaccessible areas in target-detection tasks and make such undertakings on a large scale feasible. They are a great help in search and rescue, wildfire detection or wildlife observation. However, for airborne observers, occlusion caused by vegetation adds another layer of complexity. 

\gls{acr:AOS} \cite{ref:aos-original,ref:aos-statistics} provides the technology to see through occluding foliage to what happens beneath. It is a synthetic-aperture light-field integral imaging technique that works on the basis of aerial photos. These photos are sampled over an area of several tens of meters. \gls{acr:AOS} computationally combines them to mimic a shallow-depth-of-field image that would have been taken by a camera with an aperture spanning the whole area. 

Blind sampling the synthetic aperture is time-consuming and non-optimal. Sequentially sampling with a single-camera drone following a predefined pattern \cite{ref:aos-original} may be cheap and easy to implement. However, it takes a long time to cover a large aperture. As a consequence, it is unsuitable for time-critical applications. What is more, it results in blurred images if the observed environment changes during the process. Speeding up the process by attaching a camera array to a single drone to sample multiple images in parallel \cite{ref:aos-1d-array} has other downsides. It is limited in size by physical constraints like the drone’s carrying capacity and unstable to fly. 

A swarm of individual drones, in contrast, allows for full flexibility. \gls{acr:AOS} benefits from exploiting local forest conditions, such as sparseness. A suitable objective function facilitates the use of an optimization algorithm to optimize for target visibility in the integral image. 


\gls{acr:PSO} fits drone swarms like a glove. Its concepts can be transferred 1-to-1 to them. Flocks of birds foraging for food served as inspiration for that swarm intelligence optimization algorithm. Individual birds in a flock need to, on the one hand, use their own cognition for local search. On the other hand, they also need to collaborate and exchange information on promising regions to guide others. So do particles in \gls{acr:PSO}. They combine random search with search based on their individual memories but also on shared knowledge. Moreover, \gls{acr:PSO} does not impose any restrictions such as differentiability on the objective function. Therefore it is suitable for \gls{acr:AOS}. 

This synergy has already been investigated by a previous work that presents a \gls{acr:PSO} strategy for \gls{acr:AOS} drone swarms \cite{ref:aos-pso} which we call AOS-PSO for short throughout this paper. However, their method adapts a quite standard PSO without considering the many modifications made over time to improve the PSO algorithm. In particular, it keeps its parameters fixed. They are neither reduced over time \cite{ref:pso-v-clamp} nor changed depending on the situation as \gls{acr:APSO} \cite{ref:apso} does. 

This work revisits, reviews and revises AOS-PSO. It critically examines AOS-PSO both theoretically with regard to \gls{acr:PSO} variants and its runtime artefacts. Then, it picks up one variant whose ideas are not present in this method, namely \gls{acr:APSO}. Most importantly, it combines improvements based on the analysis of AOS-PSO with \gls{acr:APSO} into a new strategy for \gls{acr:AOS} drone swarms named AOS-APSO. 

This paper continues by summarizing methods relevant to the matter in \cref{sec:bg}. That includes several \gls{acr:PSO} variants as well as PSO for AOS drone swarms. Our method is described in \cref{sec:aos-apso}. Then, \cref{sec:experiments} presents an evaluation of both the old and the new algorithm in a simulated environment along with details on their runtime behaviour. Finally, a discussion about the results, its implications and possible further improvements in \cref{sec:conclusion} concludes the work.

\section{Related Work}\label{sec:bg}

\subsection{Particle Swarm Optimization (PSO)}\label{sub:pso}

\gls{acr:PSO} \cite{ref:PSO-1,ref:PSO-2} is a stochastic, iterative optimization algorithm. It maintains a swarm, that is a set of $n$ candidate solutions called `particles'. Each particle $i$ has a position in search space $x_i(t)$ and a velocity $v_i(t)$. They are both randomly initialized for $t = 0$. The particles then move through the search space, guided by three forces as shown in the update formulas given in \cref{eq:pso-update-a,eq:pso-update-b}. 
%
$p_i(t)$ is a particle's personal best, that is the best position in terms of fitness value that the particle has seen in history. $g(t)$ denotes the global best, that is the best position over all particles.

\begin{gather}
    \label{eq:pso-update-a}
    v_i(t+1) = 
    \underbrace{\omega v_i(t)}_\text{inertia t.} + 
    \underbrace{c_{c} R_{c} \big( {p_i(t)} - x_i(t) \big)}_\text{cognitive term} + 
    \underbrace{c_{s} R_{s} \big( g(t) - x_i(t) \big)}_\text{social term}
    \\
    \label{eq:pso-update-b}
    x_i(t+1) = x_i(t) + v_i(t+1)
\end{gather}

The inertia weight $\omega$, the cognitive factor $c_c$ and the social factor $c_s$ are hyperparameters. In the original version of \gls{acr:PSO} we have $c_c = c_s = 2$ and $\omega = 1$. In fact, the inertia weight was added only in a later version of \gls{acr:PSO} \cite{ref:PSO-inertia} but has since then been considered standard 
\cite{ref:PSO-history}. 

The update is randomized by the two random variables $R_c$ and $R_s$. There is ambiguity in the literature as to whether they are scalars or vectors \cite{ref:PSO-biased}. However, it is generally assumed that random numbers are drawn uniformly in $[0,1)$.

\subsection{Charged PSO (CPSO)}\label{sub:cpso}

To improve search in dynamic environments, \gls{acr:CPSO} \cite{ref:cpso-1, ref:cpso-2} introduces repelling forces between particles to the velocity update as given in \cref{eq:cpso}. The repulsion term is inspired by Columb's Law from electrostatics. The charge of a particle $Q_i$ is a hyperparameter. To safeguard against the pole and to limit the range of influence, the summand set to zero if the distance between two particles $d_{ij}$ is zero or above a tuneable threshold $d_\text{max}$. 

\begin{gather}
    \label{eq:cpso}
    v_i(t + 1) = \ldots + \sum_{j \neq i} 
    \mathbb{1} \left( 0 < d_{ij} < d_\text{max} \right) \frac{Q_iQ_j}{||d_{ij}||^2} \frac{d_{ij}}{||d_{ij}||}
    \\ 
    d_{ij} = x_i(t) - x_j(t)
\end{gather}

The repulsion force prevents the swarm from collapsing to a single point. \gls{acr:CPSO} thus maintains swarm diversity even after convergence to an optimum. Since particles are more widely distributed, they adapt more quickly to new optima after environmental changes. 

\subsection{Adaptive PSO (APSO)}\label{sub:apso}

\gls{acr:APSO} \cite{ref:apso} dynamically adjusts \gls{acr:PSO}'s hyperparameters over time based on the swarm's configuration. Its authors observed that in the early stage of an optimization process particles are spread out through the search space. As time goes on the swarm gradually collapses to a single point. They also address dynamic environments and note that particles usually crowd around the global best. However, when the environment changes drastically, a particle that is not necessarily in the center of the swarm becomes the new global best one. 

Based on these findings, \gls{acr:APSO} employs an evolutionary factor, given in \cref{eq:aspo-ef}, that helps classify the current state of an optimization process into `exploration', `exploitation', `convergence' and `jumping out'. It is an inverse measure of the best position's centrality within the swarm. It starts at medium to high values in an exploration state and goes down to low values when converging. Jumping out happens when the evolutionary factor reaches its highest values. Unfortunately, the evolutionary factor fluctuates at a large scale. Therefore, the authors of \gls{acr:APSO} suggest to also consider the previous state for final classification so that the states are traversed in the intended order. 


\begin{gather}
    \label{eq:aspo-ef}
    e_f = \frac{d_{\text{best}} - \text{min}\left\{d_i\right\}}{\text{max}\left\{d_i\right\} - \text{min}\left\{d_i\right\}}
    \\
    d_i = \frac{1}{n-1} \sum_{j \neq i} || x_i(t) - x_j(t) ||
\end{gather}

Depending on the state, the cognitive and social factors are increased or decreased by an acceleration rate $\delta$ fixed in $[0.05, 0.1]$. Subsequently, the factors are clamped to the interval [1.5, 2.5] and normalized such that their sum does not exceed 4. Additionally, the inertia weight is adapted based on the evolutionary factor according to \cref{eq:apso-w}. 

\begin{equation}\label{eq:apso-w}
    \omega(e_f) = \frac{1}{1 + 1.5 e^{-2.6e_f}}
\end{equation}

The following summarizes what is done in which evolutionary state.

\begin{itemize}
    \item During \textbf{exploration}, the swarm's cognitive capabilities get increased while social attraction gets decreased. 
    \item To \textbf{exploit} what others have found but not nail down on it, the parameters are changed as in exploration but only with half the acceleration rate. 
    \item In a \textbf{convergence} state both factors get increased slightly to, on the one hand, guide particles to the found optimum. On the other hand, it is crucial to give the swarm the opportunity to overcome a local optimum and explore other regions. As an additional measure, the particle with the worst fitness value gets randomized at every convergence state. 
    \item When \textbf{jumping out}, the cognitive factor gets decreased and the social one increased to make the swam follow the new best position as fast as possible. 
\end{itemize}

\subsection{Rotation-Invariant PSO}\label{sub:rng-rot-inv}

\begin{figure}
    \centering
    \includegraphics[width=\linewidth]{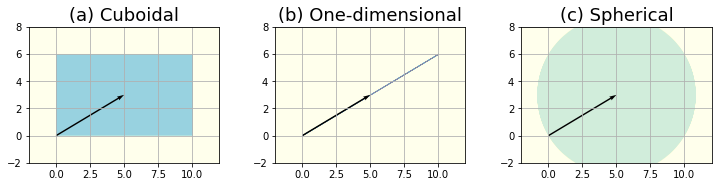}
    \caption{Supported regions of different strategies to randomize the forces in \gls{acr:PSO}.}
    \label{fig:sampling-regions}
\end{figure}

Not all strategies to generate the random components lead to rotational invariant \gls{acr:PSO} algorithms. The most commonly used way is to generate random numbers separately for each dimension of the search space. This effectively samples from a cuboid region that is parallel to the coordinate axes as depicted in \cref{fig:sampling-regions} (a). This variant is biased towards the axes; it solves problems where the solution lies on one or more axes much faster \cite{ref:PSO-biased}. The second interpretation of standard \gls{acr:PSO} is to use a random scalar as in \cref{fig:sampling-regions} (b). This is rotationally invariant, however, it restricts randomization to a single dimension. 

To solve both of these issues, \cite{ref:PSO-sphere} proposes to sample from a spherical region as given in \cref{eq:pso-sphere} for the cognitive term and analogous for the social term (\cref{fig:sampling-regions} (c)). $U_c$ is a random vector, $R_c$ a random scalar drawn uniformly in $[0, 1)$ and $m$ a positive integer. The spherical sampling is constructed such that its support contains the support of the cuboidal sampling \gls{acr:PSO}. Taking $R_c$ to the power of a positive integer $m$ compresses the distribution of radii towards smaller ones. $m = 3$ works well for most use cases \cite{ref:PSO-sphere}. 

\begin{equation}\label{eq:pso-sphere}
    c_c \Big( \underbrace{R_c^m ||p_i(t) - x_i(t)|| \frac{U_c}{||U_c||} + p_i(t)}_{\text{random point around} \ p_i(t)} - x_i(t) \Big)
\end{equation}

\subsection{Airborne Optical Sectioning (AOS)}\label{sub:aos}

\begin{figure}
    \centering
    \includegraphics[width=0.8\linewidth]{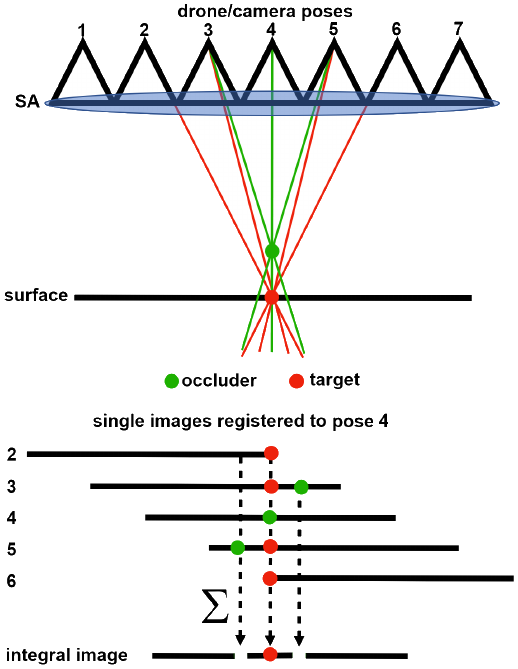}
    \caption{Schematic diagram of \gls{acr:AOS}. The upper part depicts the process of capturing the light field. The lower one illustrates the integration of individual projected images. Image by Amala Arokia Nathan et al. \cite{ref:aos-pso}, used under \href{http://creativecommons.org/licenses/by/4.0/}{CC BY 4.0}. Layout and labeling are slightly modified.}
    \label{fig:aos-schematic}
\end{figure}

\gls{acr:AOS} \cite{ref:aos-original} enables us to optically slice through an obscured volume. Essentially, it captures the light field of a scene 
and projects it back onto a virtual camera with a large aperture to produce a shallow-depth-of-field image. It obtains the light field from several sample images captured at various positions within the area of the synthetic aperture along with their pose information. The individual images come from conventional narrow-aperture cameras and therefore have a wide depth-of-field. These images are then computationally projected onto a freely chosen synthetic focal plane and integrated. Points on the focal plane align in the integral image whereas out-of-focus occluders spread out and vanish. 

Theoretical analysis from a statistical perspective on \gls{acr:AOS} \cite{ref:aos-statistics} reveals that the smallest optimal distance between sample position is the one that results in an image disparity equal to the projected occluder size. Sampling distances smaller than that result in less visibility gain; all larger ones are optimal, too (and are therefore a waste of time and drone's battery in real-world applications). Furthermore, that analysis finds that the visibility gain saturates at a high rate with the number of samples included in the integral image. A number of 10 samples already results in a visibility gain of over 95\% of the maximum possible. Combining both minimal optimal sample density and acceptable sample number leads to a well-defined recommended aperture size. 

\subsection{PSO for AOS drone swarms (AOS-PSO)}\label{sub:aos-pso}

Despite the perfect analogy between PSO's particles and physical drones, the standard PSO algorithm needs to be adjusted to meet the constraints that AOS places on the sample's positions. AOS-PSO \cite{ref:aos-pso} essentially differs from PSO in two ways best described as the `no history' and `coherent swarm' paradigm. This section reviews the assumptions that relate \gls{acr:AOS}' requirements to these paradigms and their implications on AOS-PSO. It also connects this method to other PSO variants.

For reference, the full update rule is outlined in \cref{alg:pso-for-aos}. Note that there are two different update strategies, a `default' one (if-branch) and a `\gls{acr:PSO}-like' one. 
The function $f$ abstracts the generation of an integral image from the given perspective and determines the target visibility in that image. 
It is crucial to note that the objective function depends not only on the point it is evaluated at but also on the images from other drones and previous ones, indicated by its second parameter $I(t)$. 

\begin{algorithm}
    \caption{PSO update for AOS drone swarms}
    \label{alg:pso-for-aos}
    \begin{algorithmic}[1]
        \State collect images from time step t to obtain $I(t)$
        \State $\text{visibility} = \text{max}_i \Big(f\big(x_i(t), \ I(t)\big)\Big)$
        \If {visibility < threshold}
            \State $l_i(t) = c_4 \left( i - \frac{n - 1}{2} \right) \frac{w}{||w||} \ \text{with} \ w \perp SD$ 
            \State $v_i(t+1) = c_3 SD + c_5 \big(l_i(t) - x_i(t) \big)$
            \State $x_i(t+1) = x_i(t) + v_i(t+1)$
        \Else
            \State $g(t) = \text{argmax}_{x_i(t)} \Big( f \big( x_i(t), \ I(t) \big) \Big)$
            \State \label{line:aos-pso-velocity-update} $v_i(t+1) = c_c {\frac{U}{||U||}} + c_s {\frac{g(t) - x_i(t)}{||g(t) - x_i(t)||}}$
            \State \label{line:scattering} $x_i(t+1) = \text{Scatter} \big( x_i(t) + v_i(t+1), c_4 \big)$
            \State \label{line:updating-SD} update default scanning direction $SD$
        \EndIf
    \end{algorithmic}
\end{algorithm}

\subsubsection{The `no history' paradigm}\label{sub-sub:no-history}

The first paradigm limits the \gls{acr:PSO}-like update rule to not use any history but the current time step. The global best position is chosen amongst the latest set of positions, leading to the notion of a leader drone per time step. What is more, AOS-PSO does not include any form of a particle's individual cognitive memory. 
However, the same constraint isn't imposed on the objective function. It still uses samples from previous time steps in the integral image though under the condition that they improve visibility. 

This paradigm is partially based on the assumption that keeping historical data would harm searching a dynamic environment that is highly affected by randomness. To the other part, this is due to the minimal sampling distance required for a most effective use of \gls{acr:AOS}. Since the integral image might still encompass an image taken at an old position, the authors of AOS-PSO deem any attraction by a previous position counterproductive. 

\subsubsection{The `coherent swarm' paradigm}\label{sub-sub:coherent-swarm}

The second paradigm relates to the recommended synthetic aperture size explained in \ref{sub:aos}. Particles must not be distributed across the search space arbitrarily. What is more, setting up the drones to start randomly distributed over the entire search space wastes valuable resources like time and battery in a physical application. Thus the drones start in a default scanning pattern: assembled on a line and scanning in a pre-specified direction perpendicular to that line at regular intervals. This pattern is also used whenever visibility drops below a threshold, hence the `default' branch. The scanning direction is regularly updated to point towards the location where the target was last spotted as indicated on \cref{line:updating-SD}. The default pattern is furthermore intended to replace the inertia term of standard \gls{acr:PSO}.

In the PSO-like branch, attraction by a leader ensures a coherent swarm. Therefore, this force must be stronger than any other one, meaning that \cref{eq:aos-pso-parameter-constraint-1} must hold. When the swarm has found an optimal spot and the environment does not change, the entire swarm should center around the leader which is above the optimum. Furthermore, a second constraint given in \cref{eq:aos-pso-parameter-constraint-2} where $c_4$ is the minimal optimal sample distance is intended to keep the final configuration upright. 

\begin{gather}
    \label{eq:aos-pso-parameter-constraint-1}
    c_s > c_c
    \\
    \label{eq:aos-pso-parameter-constraint-2}
    c_s + c_c \leq c_4
\end{gather}

\subsubsection{Relations to other PSO variants}\label{sub-sub:other-variants-relations}

The PSO-like velocity update on \cref{line:aos-pso-velocity-update} deserves closer examination. Due to the `no history' paradigm, the cognitive term is entirely random. To compensate, random chance is banished from the social one at all. Interestingly, the formula resembles the analogy for the social term of \cref{eq:pso-sphere} when choosing $c_c$ and $c_s$ according to \cref{eq:aos-pso-rot-inv-correspondence-1,eq:aos-pso-rot-inv-correspondence-2}. Also note that the latter term on \cref{line:aos-pso-velocity-update} vanishes for the leading drone. 

\begin{gather}
    \label{eq:aos-pso-rot-inv-correspondence-1}
    c_c = \hat{c}_s R_s^m ||x_i(t) - g(t)||
    \\
    \label{eq:aos-pso-rot-inv-correspondence-2}
    c_s = \hat{c}_s ||x_i(t) - g(t)||
\end{gather}

The Rutherford scattering implemented on \cref{line:scattering} serves to enforce a minimal distance between drones. This is required by both AOS and physical operation of drones. The formula is similar to the one used in \gls{acr:CPSO} but without the distance cutoff. What is more, AOS-PSO calculates the forces based on the new positions and applies them several times in a loop before the positions are finally sent to the drones. In contrast, \gls{acr:CPSO} calculates them once as part of the velocity update based on the old positions.

\section{Adaptive PSO strategy for AOS drone swarms (AOS-APSO)}\label{sec:aos-apso}

Our method integrates \gls{acr:APSO} into \gls{acr:PSO} for \gls{acr:AOS} drone swarms. 
It is not entirely straightforward how to do so. Both \gls{acr:APSO}'s parameter adaption and AOS-PSO need a little tweaking to work with each other efficiently. The following section describes the changes made to both to arrive at our AOS-APSO method.

\subsection{Adjustments to APSO}

\subsubsection{Evolutionary factor to state mapping}

\begin{table}
    \centering
    \begin{tabularx}{\linewidth}{|c|X|}
        \hline
        $e_f$ & evolutionary state \\
        \hline
        $< 0.025$ & convergence \\
        $[0.025, 0.1)$ & \algorithmicif \ {previous state $\in$ \{exploitation, exploration\}} \\
        & \algorithmicthen \ {exploitation} \algorithmicelse \ {convergence} \algorithmicend\algorithmicif \\
        $[0.1, 0.15)$ & exploitation \\
        $[0.15, 0.25)$ & \algorithmicif \ {previous state $\in$ \{exploration, jumping out\}} \\
        & \algorithmicthen \ {exploration} \algorithmicelse \ {exploitation} \algorithmicend\algorithmicif \\
        $[0.25, 0.5)$ & exploration \\
        $[0.5, 0.6)$ & \algorithmicif \ {previous state $\in$ \{jumping out, convergence\}} \\
        & \algorithmicthen \ {jumping out} \algorithmicelse \ {exploration} \algorithmicend\algorithmicif \\
        $\ge 0.6$ & jumping out \\
        \hline
    \end{tabularx}
    \caption{Mapping evolutionary factor to state in AOS-APSO}
    \label{tab:map-ef-to-state}
\end{table}

AOS-APSO uses lower bounds for mapping an evolutionary factor to a state compared to \gls{acr:APSO} based on early experiments with an untuned variant of AOS-APSO. The full mapping algorithm is outlined in \cref{tab:map-ef-to-state}.

\subsubsection{Modified parameter adaption}

For adapting the social and cognitive parameters, 
we introduce two auxiliary variables $\hat{c}_s(t)$ and $\hat{c}_c(t)$. They 
take on values in the open interval from 0 to 1. These fractional factors are translated to social and cognitive weight respectively by \cref{eq:aos-apso-soc-translation,eq:aos-apso-cog-translation}. The idea behind this is to respect the two constraints from PSO for AOS given in \cref{eq:aos-pso-parameter-constraint-1,eq:aos-pso-parameter-constraint-2} automatically. To fully support the second one, however, AOS-APSO has to normalize the fractional factors when $\hat{c}_s(t) \big(1 + \hat{c}_c(t)\big) > 1$. 

\begin{gather}
    \label{eq:aos-apso-soc-translation}
    c_s(t) = c_4 \hat{c}_s(t)
    \\
    \label{eq:aos-apso-cog-translation}
    c_c(t) = c_s c_c(t)
\end{gather}

Instead of an acceleration rate, AOS-APSO uses a deceleration factor $\gamma$. Whenever \gls{acr:APSO} increases a parameter, AOS-APSO uses \cref{eq:aos-apso-increase} and for decreasing it uses \cref{eq:aos-apso-decrease}. Here, $s/c$ means either $s$ or $c$. The direction of the parameter updates and the fact that slight updates use half of the deceleration factor stay unchanged. 

\begin{gather}
    \label{eq:aos-apso-increase}
    c_{s/c}(t+1) = 1 - \gamma ( 1 - c_{s/c}(t))
    \\
    \label{eq:aos-apso-decrease}
    c_{s/c}(t+1) = \gamma c_{s/c}(t)
\end{gather}

\subsubsection{Unused \gls{acr:APSO} concepts}

Furthermore, other concepts from \gls{acr:APSO} are not used in AOS-APSO. The randomization of the worst particle in a convergence state is inappropriate because it violates the `coherent swarm' paradigm. Also, inertia weight adaption is left out since PSO for AOS has no inertia term and weight as such. 

\subsection{Modifications to AOS-PSO}

\subsubsection{Parameter adaption and reset}

Parameter adaption concerns only the PSO-like branch. As a consequence, the question arises as to 
how to deal with the default branch. Our AOS-APSO simply resets the two factors to their initial values there. It thus creates a situation similar to the start whenever the target is lost. We chose the initial values according to \cref{eq:aso-apso-reset} as they approximately correspond to the fixed social and cognitive factors used in previous experiments with AOS-PSO \cite{ref:aos-pso}. The exact equivalence is given by \cref{eq:aos-apso-exact-equivalent} for reference. 

\begin{gather}
    \label{eq:aso-apso-reset}
    \hat{c}_s(t) = \hat{c}_c(t) = 0.5
    \\
    \label{eq:aos-apso-exact-equivalent}
    \hat{c}_s(t) = 0.47619\ldots, \quad \hat{c}_c(t) = 0.5
\end{gather}

\subsubsection{Leader-stabilized Rutherford scattering}

Besides integrating parameter adaption, our method does two more changes to AOS-PSO. The first one, termed `leader stabilization' is a modification to Rutherford scattering. It draws on the observation that in AOS-PSO a leader in the outskirts of a swarm gets pushed away from the rest by repelling forces. The other particles therefore almost never succeed to catch up with the leader, as argued in \cref{sub-sub:results-leader-stabilization}. To remedy this, AOS-APSO uses a leader-stabilized Rutherford scattering where the leading particle does not experience any repulsive forces. Instead, to ensure sufficient repulsion, the repelling forces other particles experience from the leader are doubled. 

\subsubsection{Projection-based default line positions}

The second modification concerns the default scanning pattern. In particular it deals with the order in which particles are assembled on the default line, that is the line perpendicular to the default scanning direction. AOS-PSO orders them by particle index. For AOS-APSO, we decided to position them in order of the projections of the drones' current positions onto the default line. This minimizes the distance the drones have to travel to reach their line positions. Furthermore, it avoids crossings of the drone's paths. This, however, is a minor modification only relevant to applications in the real world. It does neither affect AOS-PSO's PSO-like update rule nor is it relevant to \gls{acr:APSO} or improving visibility.

\section{Results}\label{sec:experiments} 

This section shows four variants of PSO for AOS drone swarms in action. The first one is AOS-PSO as originally presented \cite{ref:aos-pso} except that it uses projection-based default line positions (what is only relevant to the rare cases where the target is lost). Second, we have an AOS-PSO that additionally uses leader-stabilized Rutherford scattering to test its effects in isolation. Furthermore, `tuned AOS-APSO' refers to the adaptive PSO strategy for AOS described in \cref{sec:aos-apso}. Last but not least, `untuned AOS-APSO' uses parameter adaption and evolutionary factor to state mapping exactly as proposed for \gls{acr:APSO} but is otherwise like tuned AOS-APSO. 

The remainder of this section first familiarizes with the experimental setup. Then, it compares overall performance of the original and adaptive versions of PSO for AOS and all above-described intermediate ones. Lastly, it discusses details on individual features of this variants and the individual modifications.

\subsection{Experimental setup}\label{sub:setup}

The algorithms were tested on a procedural forest model. The scenario is simplistic: one hectare of planar ground and a static person at a fixed distance to the swarm's starting position as a target. We fixed three different initializations to test all algorithms under same conditions. An initialization encompasses the target's position, the swarm's initial position, and the seed for \gls{acr:PSO}'s pseudo random number generation. Furthermore, all runs were set to stop at a fixed limit of 20 seconds of simulated time. 

Iterations do not happen at regular intervals in the simulation. Instead, each velocity serves as a distance to be covered. Drones travel these distances at a constant speed of 10 m/s. The longer the distance, the longer the simulated time required for the corresponding time step. This allows to interpret the results in a way that is more meaningful for real-world applications than iteration counts. The effects of processing time as well as drone acceleration and deceleration are not taken into account.

\subsection{Evaluation}\label{sub:performance}

\begin{figure}
    \centering
    \makebox[\linewidth]{\includegraphics[width=1.1\linewidth]{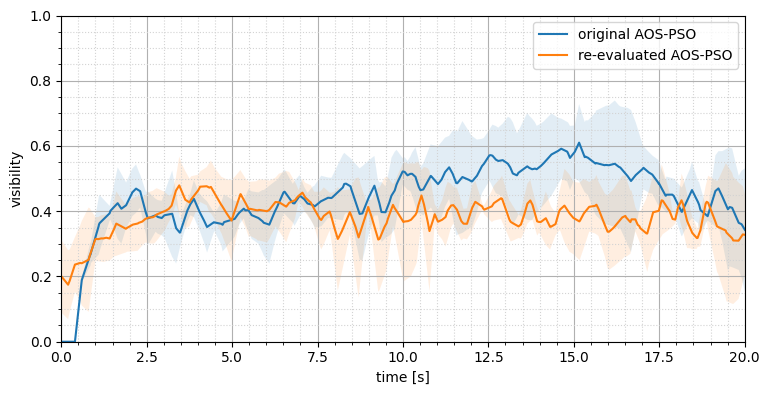}}
    \caption{Results on AOS-PSO reported by \cite{ref:aos-pso} vs. our own. Averaged linearly interpolated visibility-time curves over three experiments with 95\% confidence intervals.}
    \label{fig:reevaluation-aos-pso}
\end{figure}

\begin{figure}
    \centering
    \makebox[\linewidth]{\includegraphics[width=1.1\linewidth]{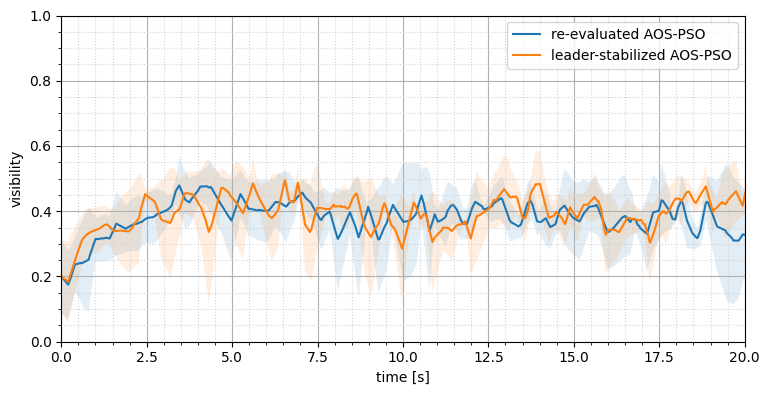}}
    \caption{Effects of leader stabilisation on AOS-PSO. Averaged linearly interpolated visibility-time curves over three experiments with 95\% confidence intervals.}
    \label{fig:results-leader-stabilization}
\end{figure}

\begin{figure*}
    \centering
    \includegraphics[width=0.7\linewidth]{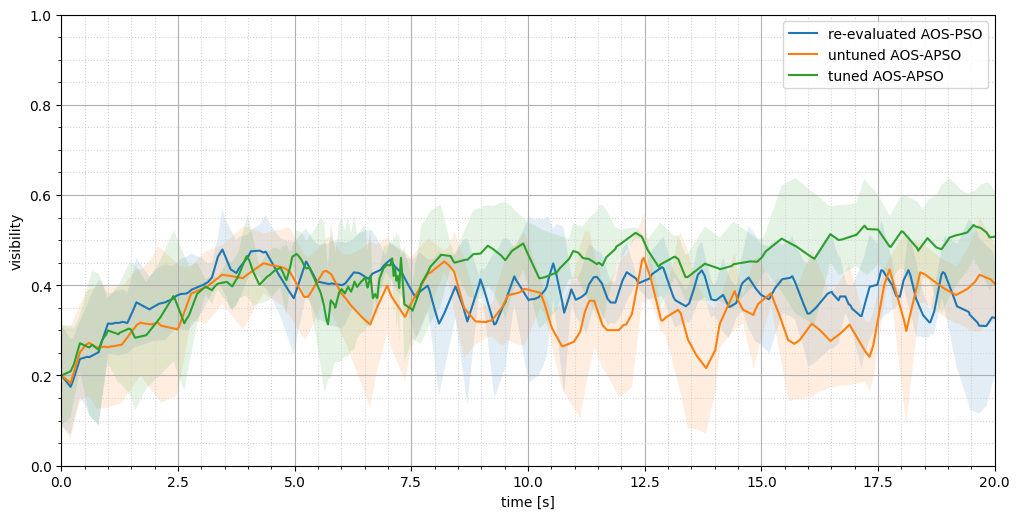}
    \caption{Adaptive vs. non-adaptive PSO for AOS drone swarms. Averaged linearly interpolated visibility-time curves over three experiments with 95\% confidence intervals.}
    \label{fig:results-aos-apso}
\end{figure*}

The graphs in this section report the mean visibility over time of the different algorithms. The results from the three individual experiments per algorithm were linearly interpolated before averaging them. Additionally, the graphs include the 95\% confidence intervals for these averages as areas of equal color but semitransparent. 
The visibility axis is normalized by the maximum visibility possible when the target is observed unoccluded. The time axis indicates simulation time rather than iteration count or processing time. 

We decided to reevaluate the original AOS-PSO under the same three initializations to keep the results comparable. The results are contrasted with the ones originally reported \cite{ref:aos-pso} in \cref{fig:reevaluation-aos-pso}. This graph already turned out to be insightful. The first phase where visibility improves quickly to a value around 0.4 is present in both experiments, although it takes longer in the reevaluated runs. This is probably due to the different distances that the swarm's starting positions have to the targets in these two experiment settings. What is most interesting is that the original runs continue to rise to a maximum of 0.6 before falling down again, while our runs stagnate at around 0.4.

In contrast to the large gap between experiments shown in \cref{fig:reevaluation-aos-pso}, the graphs in both \cref{fig:results-leader-stabilization} and \cref{fig:results-aos-apso} barely differ at a significant level. At least this means that the leader-stabilized version of AOS-PSO (orange in \cref{fig:results-leader-stabilization}) is no worse than the unmodified one. Moreover, the tuned AOS-APSO (green in \cref{fig:results-aos-apso}) shows a tendency to improve over the reevaluated AOS-PSO over time. Its untuned version (orange in \cref{fig:results-aos-apso}), in contrast, tends to underperform the baseline.

\subsection{Details}\label{sub:result-details}

The following covers details on individual experiments that illustrate particular features of the algorithms being tested. It begins with an investigation of tuned AOS-APSO and then goes on to findings that support the modifications made to AOS-PSO which do not draw on ideas from \gls{acr:APSO}. 
It does so mainly by showing trajectories of a swarm's particles in search space. The temporal development is coded by color in these graphs, from darker to lighter ones. The leader particles have a different color gradient and a star shape instead of a circular one. Note that the target's position does not correspond to the point of highest visibility. In fact, the objective function assigns higher values the more oblique the target is viewed. 
For associated visibility and other values and to put the examples shown in a larger context, refer to \cref{apx:individual-experiments}.

\subsubsection{A flaw of tuned AOS-APSO}\label{sub-sub:tuned-aos-apso-flaw}

\begin{figure}
    \centering
    \makebox[\linewidth]{\includegraphics[width=1.1\linewidth]{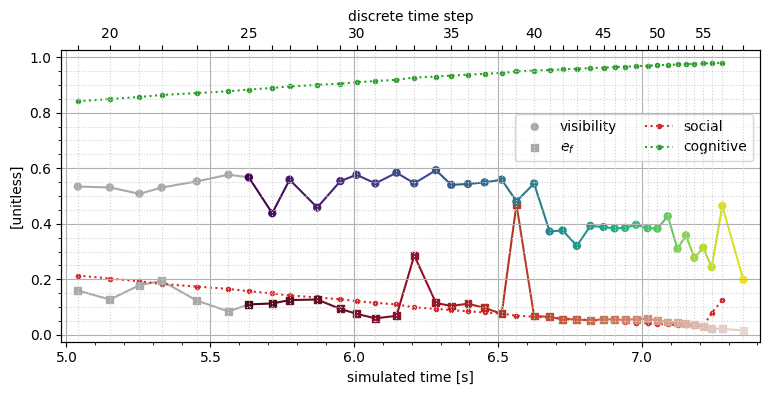}}
    \caption{Experiment 2 from tuned AOS-PSO (clipped). Colors per time step correspond to \cref{fig:tuned-aos-apso-exp-2-explosion}.} 
    \label{fig:tuned-aos-apso-exp-2}
\end{figure}

It is worth noting the unique feature of tuned AOS-APSO between 5 and 7.5 seconds in \cref{fig:results-aos-apso}. It uncovers a particular flaw of this algorithm, namely an imbalance between repulsive and attractive forces on particles. The artefact stems from the experiment 2 which is way more granular in this region than anywhere else or any other curve. \Cref{fig:tuned-aos-apso-exp-2} shows the relevant section of its visibility curve along with evolutionary, social and cognitive factor. Most interestingly, visibility drops at some point in this interval. 

\begin{figure}
    \centering
    \includegraphics[width=\linewidth]{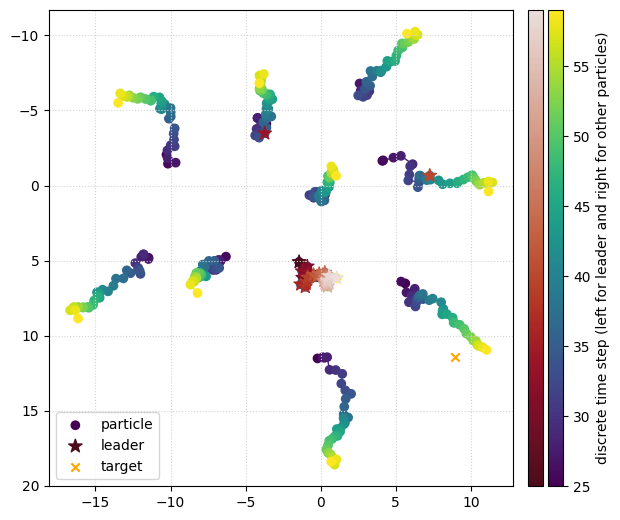}
    \caption{Particle trajectories in search space from experiment 2 of tuned AOS-APSO. The particles flee in an almost straight path from the leader.}
    \label{fig:tuned-aos-apso-exp-2-explosion}
\end{figure}

Investigating the precise drone configurations at these time steps, depicted in \cref{fig:tuned-aos-apso-exp-2-explosion}, shows that the drones move further and further away from the center. Since tuned AOS-APSO is designed such that the cognitive term's magnitude is always a fraction from the social one's, it is unlikely that this is due to random chance. Note that the granularity implies that the largest movement and hence the maximum velocity $\text{max}\left( v_i(t) \right)$ at the respective $t$'s is small. 
Apparently, movements get so small that the repelling forces from Rutherford scattering dominate the force towards the leader. Eventually, drones that do not lead but still contribute a significant part to good visibility loose their good spots and visibility drops. 

\subsubsection{Leader stabilization}\label{sub-sub:results-leader-stabilization}

Plotting the evolutionary factor of the experiments on AOS-PSO over time (first column of \cref{fig:grid-of-curves-original}) immediately reveals that this algorithm does not fulfill the premise of \gls{acr:APSO}. The evolutionary factor starts low, but soon rises and then remains high most of the time except for some jumps that last for a few time steps at maximum. This is unfavorable for applying \gls{acr:APSO} to AOS-PSO, but in fact it also comes unexpected for AOS-PSO itself. 
AOS-PSO's main force is the attraction by the leader. So, one would expect it to converge to states of high leader centrality what is equivalent to low values of $e_f$. 
The more in-depth analysis presented in the following suggests that the issue lies in Rutherford scattering employed in PSO for AOS. It also shows how the leader stabilization modification presented at the end of \cref{sec:aos-apso} solves this. 

\begin{figure}
    \centering
    \includegraphics[width=\linewidth]{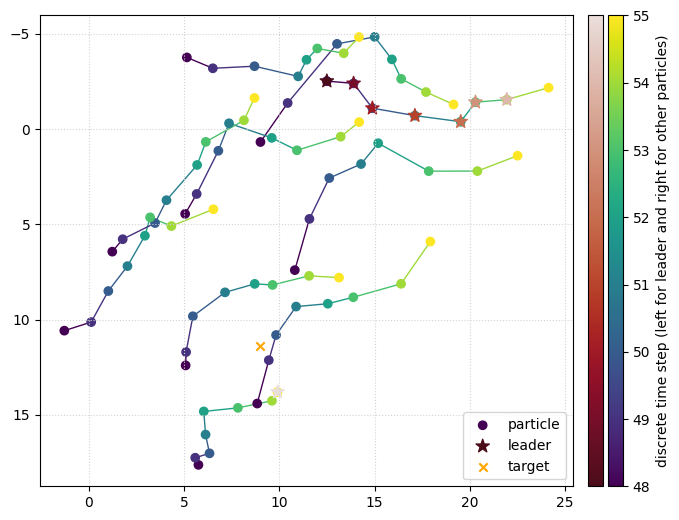}
    \caption{Particle trajectories in search space from experiment 2 of AOS-PSO. The swarm follows a leader (in the top right corner) who is pushed ahead of it.}
    \label{fig:aos-pso-exp-2-leader-pushing}
\end{figure}

In AOS-PSO, the swarm frequently exhibits a behaviour similar to the one shown in \cref{fig:aos-pso-exp-2-leader-pushing}. A particle at the outskirts of the swarm becomes the leader and the remaining ones are unable to catch up with it. This behaviour appears too consistently to have its cause in the random component of AOS-PSO. So the only possible explanation is that this happens due to the repelling forces from Rutherford scattering. 

\begin{figure}
    \centering
    \includegraphics[width=\linewidth]{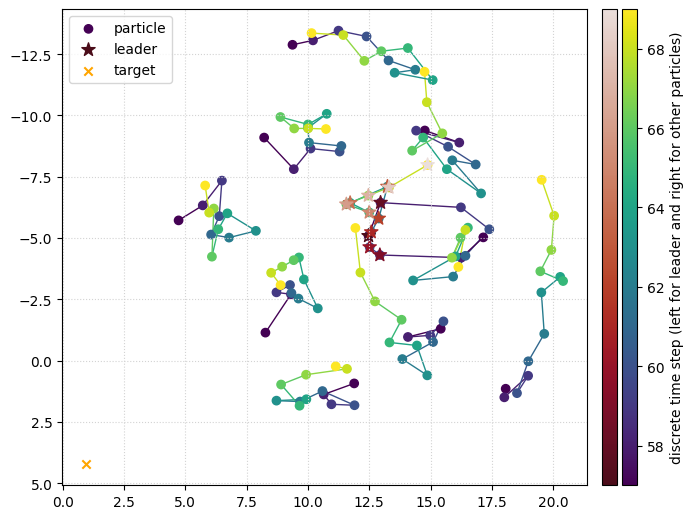}
    \caption{Particle trajectories in search space from experiment 1 of AOS-PSO. The swarm managed to center the leader and maintains this configuration.}
    \label{fig:aos-pso-exp-1-leader-swarming}
\end{figure}

There is a single clear exception to this behaviour in experiment 1, depicted in \cref{fig:aos-pso-exp-1-leader-swarming}. Here, the swarm perfectly centers the leader and all particles fluctuate only in their own local spheres. However, this is the only case from AOS-PSO where the swarm reaches this final configuration although it is intended to converge to it by both, PSO for \gls{acr:AOS} drone swarms and, according to APSO's analysis, PSO in general. 

\begin{figure}
    \centering
    \includegraphics[width=\linewidth]{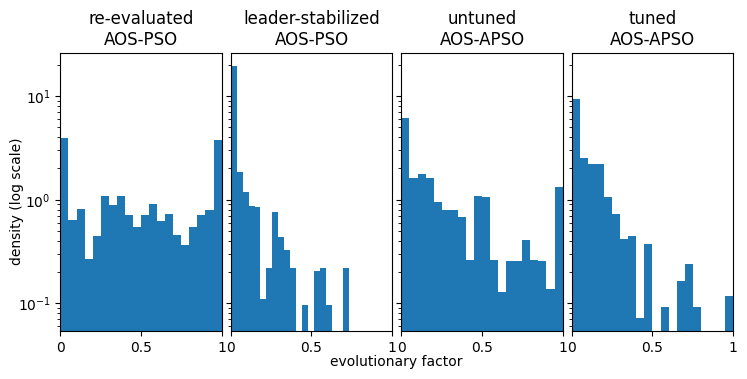}
    \caption{Distribution of evolutionary factors exhibited by the inspected methods. The leftmost one does not use leader-stabilized Rutherford scattering whereas all others do.}
    \label{fig:ef-histograms}
\end{figure}

The experiments with leader-stabilised AOS-PSO exhibit generally lower evolutionary factors. \Cref{fig:ef-histograms} summarizes the distribution of $e_f$ values of all algorithms to show this at a glance. While the original AOS-PSO has a distribution over the entire spectrum with a tendency to extremes, it is skewed to lower values for the leader-stabilized version. The distributions for the two AOS-APSO variants, that use leader stabilization, too, are less extreme but still tilted towards the left.

\subsubsection{Default line positions}\label{sub-sub:projection-based-default-line}

\begin{figure}
    \centering
    \includegraphics[width=0.7\linewidth]{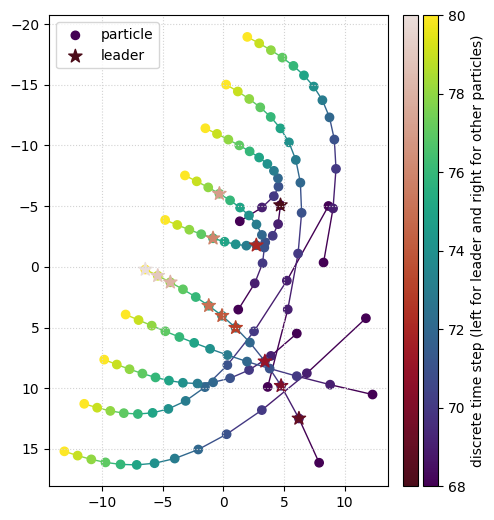}
    \caption{Particle trajectories in search space from a separate run of AOS-PSO with index-based default line positions. The particles have to travel a large distance and cross each others paths to assemble in the default scanning pattern.}
    \label{fig:separate-path-crossing}
\end{figure}

\begin{figure}
    \centering
    \includegraphics[width=0.7\linewidth]{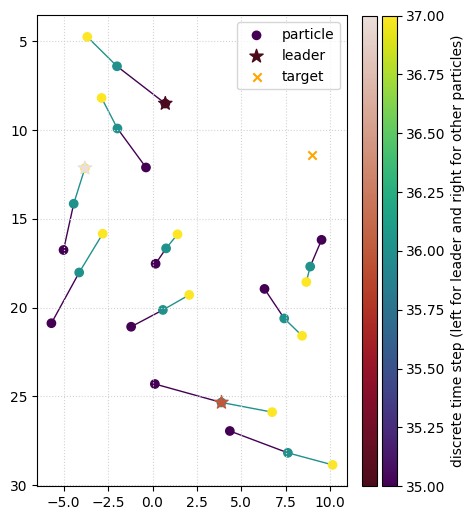}
    \caption{Particle trajectories in search space from experiment 2 of untuned AOS-APSO. Particles begin to assemble on the default line according to their projection onto it.}
    \label{fig:experiment-projection-based-line-positions}
\end{figure}

\Cref{fig:separate-path-crossing} exemplifies how default line positions by particle index lead to inefficient particle trajectories. It is taken from a separate uncontrolled experimental run. There the swarm completely lost the target. In contrast, \cref{fig:experiment-projection-based-line-positions} shows the initial stage of a swarm following a particle-positions-based default scanning pattern. Clearly, in the latter case paths are shorter, more direct and crossing-free. 

\subsubsection{Exact positions matter}\label{sub-sub:exact-positions-matter}

\begin{figure}
    \centering
    \makebox[\linewidth]{\includegraphics[width=1.1\linewidth]{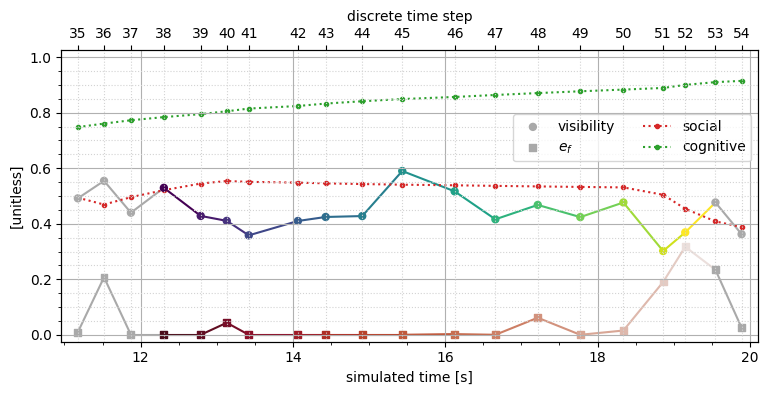}}
    \caption{Experiment 1 of tuned AOS-APSO (clipped). Colors per time step correspond to \cref{fig:vis-drop-trajectories}.}
    \label{fig:vis-drop-curves}
\end{figure}

\begin{figure}
    \centering
    \includegraphics[width=\linewidth]{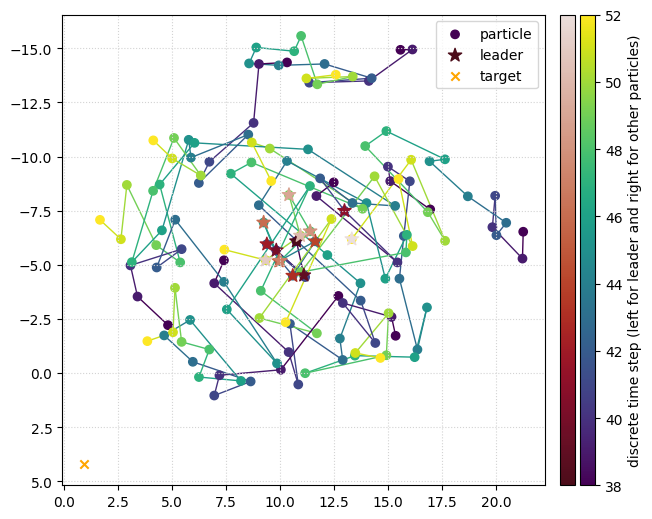}
    \caption{Particle trajectories in search space from experiment 1 of tuned AOS-APSO. The swarm keeps covering a fixed area in which particles move freely.}
    \label{fig:vis-drop-trajectories}
\end{figure}

\Cref{fig:vis-drop-curves,fig:vis-drop-trajectories} show a situation from tuned adaptive PSO for AOS where the evolutionary factor is consistently low (below 0.1) for a relatively long period of about 6 to 7 seconds of simulated time. As expected, the area covered by the swam and the approximate position of the leader remain largely unchanged. Interestingly, visibility fluctuates at a scale similar to other situations where the swarm's configuration is less constant. This indicates that within the area, the exact positions of the drones still matter.

\section{Conclusion}\label{sec:conclusion}

The overall results shown in \cref{fig:results-aos-apso} are -- though our tuned adaptive AOS-APSO improved in this experiments over the virtually original AOS-PSO on average -- neither significant nor more than marginal. What is more, none of our experimental runs exceeds or even reaches the results reported in \cite{ref:aos-pso}. This leads to the conclusion that randomness and initial conditions highly affect the results. It seems that under the given paradigms AOS-PSO is already quite close to the best we can do. 

A possible further research direction is to weaken the `no history' paradigm. A good swarm configuration is not only determined by the position and size of the synthetic aperture they form but also by the exact positions of the drones within that area. That is exactly the reason for using drone swarms for \gls{acr:AOS} in the first place. It is also indicated by the experiments presented here, particularly \cref{fig:tuned-aos-apso-exp-2-explosion,fig:vis-drop-trajectories}. The `no history' paradigm, however, prohibits any form of a particle's individual memory. As a consequence, particles sooner or later drift off their good positions. They have no means of remembering where to go back when random search in another region turns out to be fruitless. 

Moreover, the assumptions that led to this paradigm in the first place themselves are questionable. First of all, simply forgetting old positions does not prevent the swarm from sampling multiple times from the same position. On the contrary, memorizing a previous position allows a particle to actively avoid the exact same spot. This could be done for example by an appropriate probability distribution for the random components of \gls{acr:PSO} similar to the concept of \cite{ref:PSO-sphere}. Secondly, \gls{acr:PSO} literature already provides sophisticated methods to deal with (highly) dynamic environments. \gls{acr:APSO} and \gls{acr:CPSO} are only two examples. They both still use a cognitive memory for particles. 

Any future work on the adaptive PSO for AOS should definitely first deal with its open issue. Introducing narrower bounds for the fractional factors $\hat{c}_s(t)$ and $\hat{c}_c(t)$ is a possible remedy to investigate. Another possibility is to utilize \gls{acr:CPSO}'s idea to introduce a cutoff distance to repelling forces. In any case, more experiments are definitely needed to draw a final conclusion on whether AOS-APSO pays off at all. Particularly, they should include a wider range of situations with dynamics involved, e.g. by moving targets, to examine \gls{acr:APSO}'s full strength on AOS drone swarms. And again, it would be interesting to check out how a personal best memory for particles affects AOS-APSO's performance and behaviour.

\begin{acks}
Many thanks to Rakesh Amala Arokia Nathan for providing the source code and answering question about his PSO strategy for AOS drone swarms. Also, I want to acknowledge Lukas Bostelmann-Arp for sharing his speed-improved forest simulation even though it was finally not used for this work. And last but not least, thanks to my supervisor Oliver Bimber. 
\end{acks}

\bibliographystyle{ACM-Reference-Format}
\bibliography{references}

\appendix

\section{Individual Experiments}\label{apx:individual-experiments}

\begin{figure*}
    \centering
    \includegraphics[width=\linewidth]{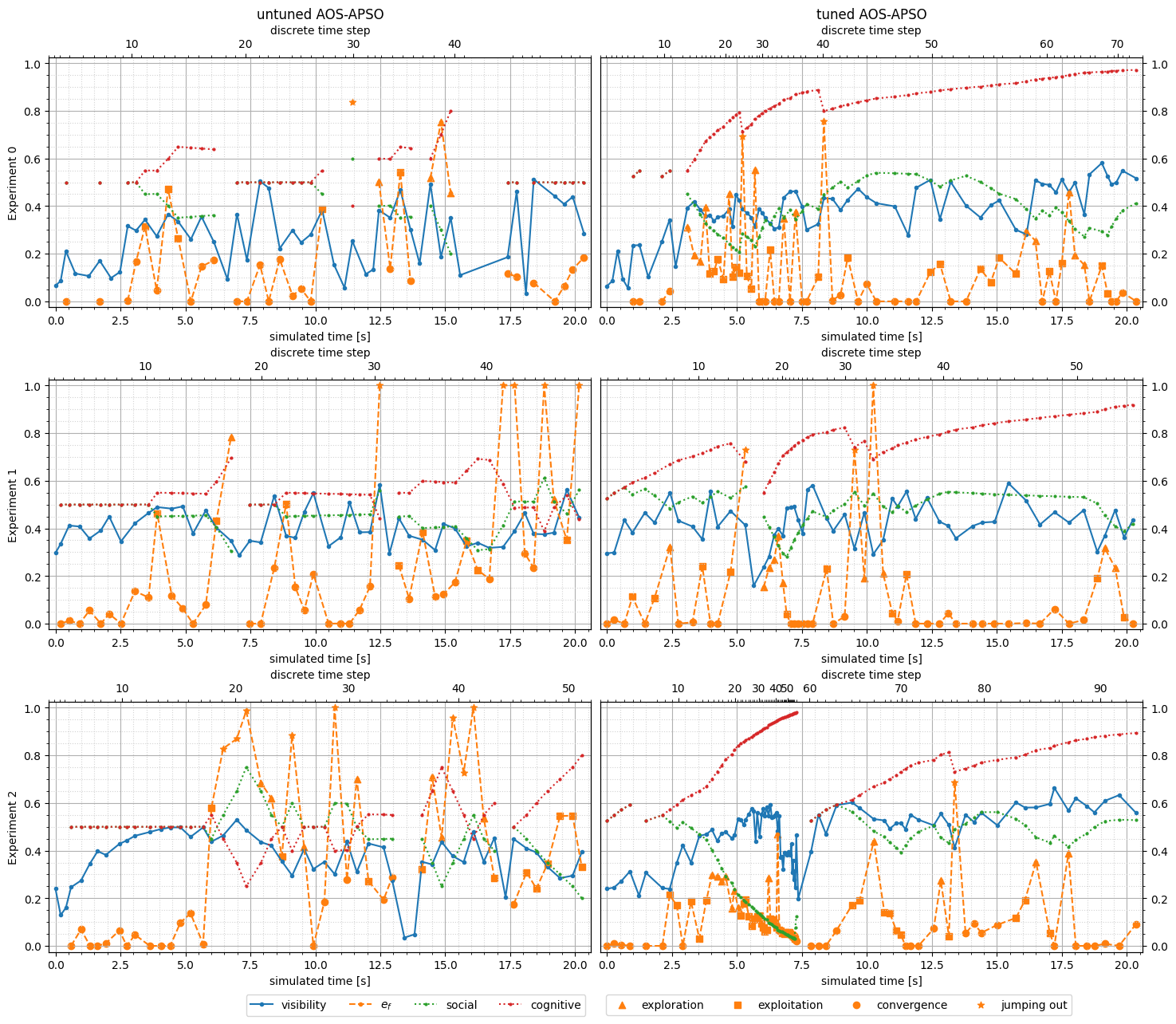}
    \caption{Individual experiments with adaptive methods. The shape of an $e_f$ marker indicates the associated evolutionary state. For untuned AOS-APSO (left column), social and cognitive factors are normalized to \gls{acr:APSO}'s clamping range [1.5, 2.5]. For tuned AOS-APSO (right column), the graphs show the fractional factors $\hat{c}_s$ and $\hat{c}_c$ instead of social and cognitive ones respectively.}
    \label{fig:grid-of-curves-adaptive}
\end{figure*}

\begin{figure*}
    \centering
    \includegraphics[width=\linewidth]{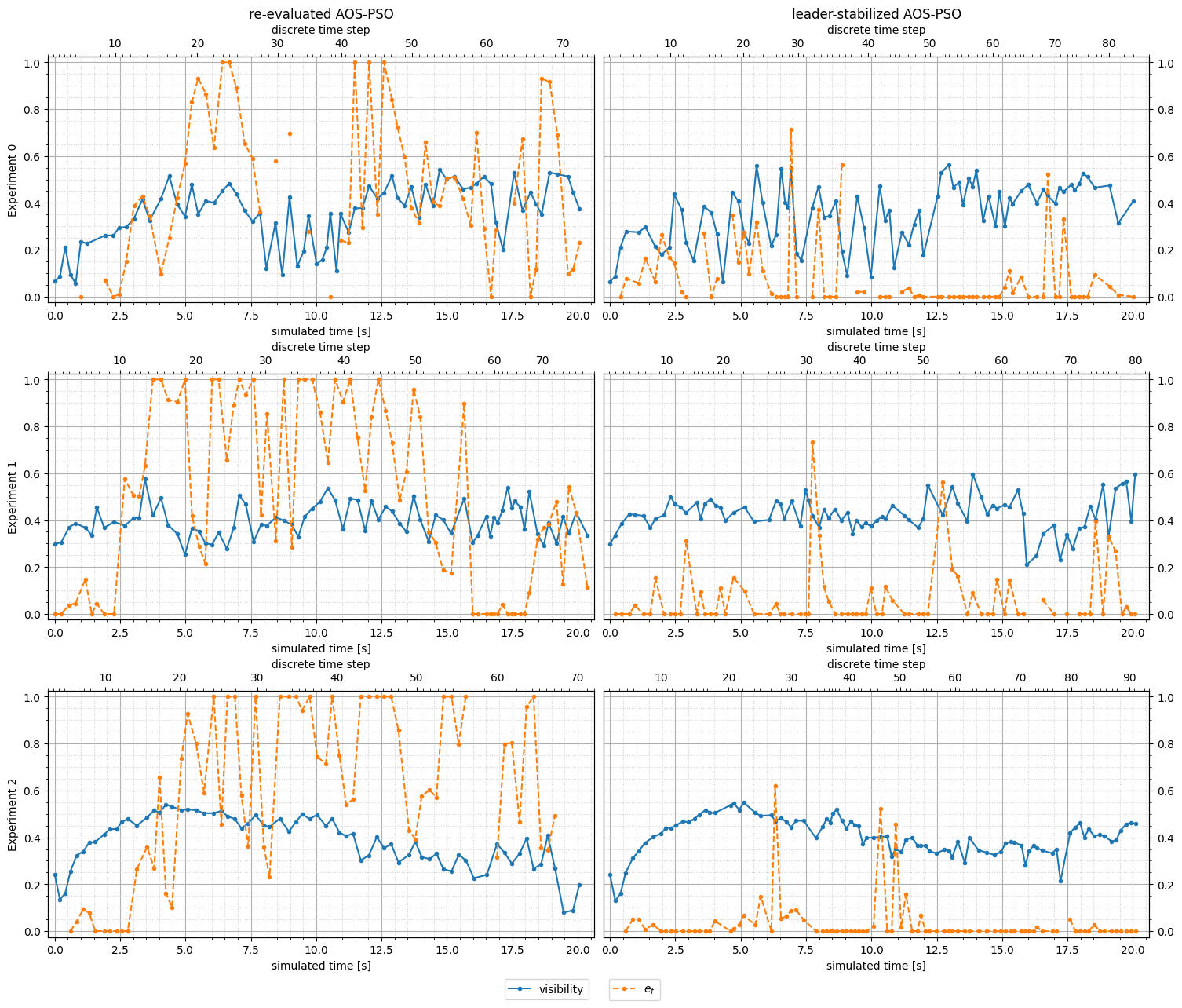}
    \caption{Individual experiments with non-adaptive methods.}
    \label{fig:grid-of-curves-original}
\end{figure*}

\Cref{fig:grid-of-curves-adaptive,fig:grid-of-curves-original} show the values of visibility and evolutionary factor over time from individual experiments with adaptive and non-adaptive algorithms respectively. For the adaptive ones, they also include evolutionary state, indicated by the shape of the $e_f$ markers, and normalized social as well as cognitive factors. 

\end{document}